\documentclass[12pt]{article}
\usepackage[english]{babel}
\usepackage{mathtext}
\usepackage[T2A]{fontenc}
\usepackage{epsfig}
\begin{document}
%\renewcommand{\figurename}{Fig.}
%\begin{titlepage}
\begin{center} {\Large {\bf Maximum Renyi entropy principle for systems
with power--law Hamiltonians}}\\
% \vspace{.4cm}
  \vspace{.2cm}
\renewcommand{\thefootnote}{\fnsymbol{footnote}}
%\vspace{.23cm}
\vspace{.5cm} \large {A.G.Bashkirov}\footnote{{\it E-mail
address}: abas@idg.chph.ras.ru}\\ \vspace{.25cm} Institute
Dynamics of Geospheres, RAS,\\ Leninskii prosp. 38 (bldg.1),
119334, Moscow, Russia\\
\end{center}
\begin{abstract}
The Renyi distribution ensuring the maximum of a Renyi entropy is
investigated for a particular case of a power--law Hamiltonian.
Both Lagrange parameters, $\alpha$ and $\beta$ can be excluded. It
is found that $\beta$ does not depend on a Renyi parameter $q$ and
can be expressed in terms of an exponent $\kappa$ of the
power--law Hamiltonian and an average energy $U$. The Renyi
entropy for the resulted Renyi distribution reaches its maximal
value at $q=1/(1+\kappa)$ that can be considered as the most
probable value of $q$ when we have no additional information on
behaviour of the stochastic process. The  Renyi distribution for
such $q$ becomes a power--law distribution with the exponent
$-(\kappa +1)$. When $q=1/(1+\kappa)+\epsilon$ ($0<\epsilon\ll 1$)
there appears a horizontal "head" part of the Renyi distribution
that precedes the power--law part. Such a picture corresponds to
observables.
\end{abstract}
 \vspace{.1cm}
PACS:  05.10.Gg, 05.20.Gg, 05.40.-a
 \setcounter{footnote}{0}

\section{Introduction}

Numerous examples of power--law distributions (PLD) are well-known
in different fields of science and human activity \cite{Mant}.
Power laws are considered \cite{Bak2} as one of signatures of
complex self-organizing systems. They are sometimes called
Zipf--Pareto law or fractal distributions. We can mention here the
Zipf--Pareto law in linguistics \cite{Zipf}, economy \cite{Mand2}
and  in the science of sciences \cite{Price}, Gutenberg-Richter
law in geophysics \cite{Gut}, PLD in critical phenomena
\cite{Baxt}, PLD of avalanche sizes in sandpile model for
granulated media \cite{Bak1} and fragment masses in the impact
fragmentation \cite{Fujiw,fragm}, etc.

Graphically, PLD is presented by a linear graph in a double
logarithmic plot of frequency or cumulative number as a function
of size. It should be noticed here that, in general, double
logarithmic plots of data from phenomena in nature or economy
often exhibit a limited linear regime preceded by a
near-horizontal "head" part and followed by a tail of significant
curvature. The latter deviation from a power--law description can
be explained by a finite-size effect. Really, for instance for the
impact fragmentation, extrapolation of the PLD to infinite
fragment masses would predict masses surpassing a mass of the
target. This effect will not be considered here.

Successful derivations of PLD with the head part are based on
Renyi or Tsallis distributions ensuring maximums of Renyi or
Tsallis entropies correspondingly (see,e. g. \cite{BaVit,Tsall}).
However, some parameters ($q$-parameter, Lagrange multiplier
$\beta$) were left to be indeterminate there.

Here, a derivation of PLD will be performed on the base of maximum
entropy principle (MEP) for the Renyi entropy. In the special case
that a Hamiltonian of the system is a power--law function of a
variable of the system the  Lagrange multiplier $\beta$ will be
excluded at all from the Renyi distribution (Sec. 2) and the
$q$-parameter will be determined uniquely with the use of
expansion of the MEP (Sec. 3).

According to the well-known maximum entropy principle (MEP)
developed by Jaynes \cite{Jaynes} for a Boltzmann-Gibbs statistics
an equilibrium distribution of probabilities $p=\{p_i\}$ must
provide maximum of the Boltzmann  information entropy $S_B$ upon
additional conditions of normalization  $\sum_i p_i=1$ and a fixed
average energy $ U=\langle H\rangle_p\equiv\sum_i H_i p _i$.

Then, the distribution $\{p_i\}$ is determined from the extremum
of the functional
 \begin{equation}
L_G(p )=- \sum_i^W\,p_i\ln p_i - \alpha_0 \sum_i^W\,p _i - \beta_0
\,\sum_i^W H_i p _i ,
\end{equation}
where $ \alpha_0 $ and $\beta_0 $  are Lagrange multipliers. Its
extremum is ensured by the Gibbs canonical distribution, in which
$\beta_0 =1/k_BT$ and $T$ is the thermodynamic temperature, so
there is no necessity to invoke the average energy $U$ for
exclusion of the Lagrange parameter $\beta_0$ from the Gibbs
distribution because of a well-established correspondence between
Gibbs thermostatistics and classical thermodynamics. On contrary,
in applications of MEP to the Renyi entropy, the Lagrange
parameter $\beta$ may depend on the $q$-parameter and its physical
meaning is not evident, so it is necessary to exclude it with the
use of the additional condition of MEP related to the fixed
average energy.

\section{MEP for the Renyi entropy}

If the Renyi entropy
\begin{equation}
S_R=\frac {k_B}{1-q}\ln \sum_i p_i^{q }
\end{equation}
is used instead of the Boltzmann entropy the equilibrium
distribution must ensure maximum of the functional
\begin{equation}
L_R(p )=\frac 1{1-q}\,\ln \sum_i^W\,p^{q } _i - \alpha \sum_i^W\,p
_i - \beta \,\sum_i^W H_i p _i,
\end{equation}
where $ \alpha $ and $\beta $  are Lagrange multipliers. Note that
$L_R(p)$ passes to $L_G(p )$ in the $q\to 1$ limit.

We equate a functional derivative of $L_R(p )$ to zero, then
\begin{equation}
\frac{\delta L_R(p )}{\delta p _i}=\frac q{1-q}\,\frac
{p_i^{q-1}}{\sum_j p_j^{q}} - \alpha -\beta H_i =0.
\end{equation}
To exclude the parameter $\alpha$ we can multiply this equation by
$p_i$ and sum up over $i$, with account of normalization condition
$\sum_i p_i=1$. Then we get
\begin{equation}
\alpha =\frac q{1-q}- \beta   U
\end{equation}
and
\begin{equation}
p _i=\left(\sum_j^W\,p^{q} _j\, (1-\beta\frac{q-1}{q} \Delta H_i)
\right)^{\frac{1}{q-1}},\,\,\Delta H_i=H_i-  U.
\end{equation}
Using once more the condition $\sum_i p_i=1$ we get $$
\sum_j^W\,p_j^q =\left(\sum_i^W (1-\beta\frac{q-1}{q}\Delta
H_i)^{\frac{1}{q-1}}\right)^{-(q-1)}$$ and, finally,
\begin{eqnarray}
p_i=p_i^{(R)} &=&Z_R^{-1}\left(1-\beta\frac{q-1}{q}\Delta
H_i\right)^{\frac{1}{q-1}}\\
Z_R^{-1}&=&\sum_i\left(1-\beta\frac{q-1}{q}\Delta
H_i\right)^{\frac{1}{q-1}}.
\end{eqnarray}
This distribution satisfies to MEP for the Renyi entropy and may
be called the Renyi distribution. When $q\to 1$ the distribution
$\{p^{(R)}_i\}$ becomes the Gibbs canonical distribution. In this
limit $\beta/q\to\beta_0=1/k_BT$. Such behaviour is not enough for
unique determination of $\beta$. Really, in general it may be an
arbitrary function $\beta(\beta_0,q)$ which becomes $\beta_0$ in
the limit $q\to 1$.

To find an explicit form of $\beta$ we return to the additional
condition of pre-assigned average energy $ U=\sum_i H_i p _i$ and
substitute there the Renyi distribution (7). For definiteness
sake, we will confine the discussion to the particular case of a
power-law dependence of the Hamiltonian on a parameter $x$
\begin{equation}
H_i=Cx_i^\kappa .
\end{equation}

If the distribution $\{p_i\}$ allows for smoothing over the range
much larger an average distance $\Delta x_i=x_i-x_{i+1}$ without
sufficient loss of information, we can pass from the discrete
variable $x_i$ to the continuous one $x$. Then the condition of a
fixed average energy becomes
\begin{equation}
Z^{-1}\int_0^\infty\,Cx^\kappa \left(1-\beta
\frac{q-1}{q}(Cx^\kappa-U)\right)^{\frac{1}{q-1}}dx =U,
\,\,\,\,H(x)=Cx^\kappa
\end{equation}
or
\begin{equation}
Z^{-1}\int_0^\infty\, C_ux^\kappa \left(1-\beta U
\frac{q-1}{q}(C_ux^\kappa-1)\right)^{\frac{1}{q-1}}dx=1,
\end{equation}
\begin{equation}
Z=\int_0^\infty\, \left(1-\beta U
\frac{q-1}{q}(C_ux^\kappa-1)\right)^{\frac{1}{q-1}}dx,\,\,\,\,C_u=\frac
C{U}.
\end{equation}
Both integrals in these equations may be calculated with the use
of a tabulated \cite{GradRyzh} integral
\begin{equation}
I=\int_0^\infty\,\frac{x^{\mu -1}dx}{(a+bx^\nu)^\lambda}=\frac
1{\nu a^\lambda} \left(\frac a{b}\right)^{\frac
\mu{\nu}}\frac{\Gamma[\,\frac \mu{\nu}\,]\Gamma[\,\lambda -\frac
\mu{\nu}\,]}{\Gamma[\,\lambda\,]}
\end{equation}
under condition of convergence
\begin{equation}
0<\frac \mu{\nu}<\lambda ,\,\,\,(\lambda >1).
\end{equation}
For the integrals in Eqs. (16) and (17) this condition becomes
\begin{equation}
0<\frac {1+\kappa}{\kappa}<\frac 1{1-q}.
\end{equation}
Then, finally, we find from Eqs. (11), (12) with the use of the
relation $\Gamma[1+z]=z\Gamma[z]$, that
\begin{equation}
\beta U =\frac 1{\kappa}\,\,\,\,{\rm for \,\,all}\,\, q.
\end{equation}
Independence of this relation on $q$ means that it is true, in
particular, for the limit case $q=1$ where the Gibbs distribution
take a place and therefore
\begin{equation}
\beta=\beta_0\equiv 1/k_BT\,\,\,{\rm for \,\,all}\,\,q.
\end{equation}
at least for power-law Hamiltonian.

When $H=p^2/2m$ (that is, $\kappa=2$) we get from (16) and (17)
that $U=\frac 1{2}k_BT$ as it should be waited for one-dimensional
ideal gas.

Besides, the Lagrange parameter $\beta$ can be eliminated from the
Renyi distribution (7) with the use of Eq. (16) and we have,
alternatively,
\begin{equation}
p^R(x|q,\kappa)=Z^{-1}\left(1- \frac{q-1}{\kappa
q}(C_ux^\kappa-1)\right)^{\frac{1}{q-1}},\,\,H(x)=Cx^\kappa
\end{equation}
or
\begin{equation}
p^R_i(q,\kappa)=Z^{-1}\left(1- \frac{q-1}{\kappa q}(C_u
x^\kappa_i-1)\right)^{\frac{1}{q-1}},\,\,\,H_i=Cx_i^\kappa .
\end{equation}
So, at least for power-law Hamiltonian, the Lagrange multiplier
$\beta$ does not depend on the Renyi parameter $q$ and coincides
with the Gibbs parameter $\beta_0=1/k_BT$, and, moreover, can be
excluded at all with the use of the relation (16).

The problem to be solved for an unique definition of the Renyi
distribution is determination of a value of the Renyi parameter
$q$. This will be the subject of the next section.

\section{The most probable value of the Renyi parameter}

An excellent example of a physical non-Gibbsian system was pointed
by Wilk and Wlodarchuk \cite{Wilk}. They took into consideration
fluctuations of both energy and temperature of a minor part of a
large equilibrium system. This is a radical difference of their
approach from the traditional Gibbs method in which temperature is
a constant characterizing the thermostat. As a result, their
approach leads \cite{BaSuk} to the Renyi distribution with the
parameter $q$ expressed via heat capacity $C_V$ of the minor
subsystem $$ q=\frac {C_V-k_B}{C_V}.$$ The approach by Wilk and
Wlodarchuk was advanced by Beck and Cohen \cite{Beck1,Beck2} who
coined it the new term "superstatistics". In the frame of
superstatistics, the parameter $q$ is defined by physical
properties of a system.

On the other side, there are many stochastic systems for which we
have no information related to a source of fluctuations. In that
cases the parameter $q$ can not be determined with the use of the
superstatistics.
\begin{figure}[t]
\begin{minipage}{.50\linewidth}
 \centering\epsfig{figure=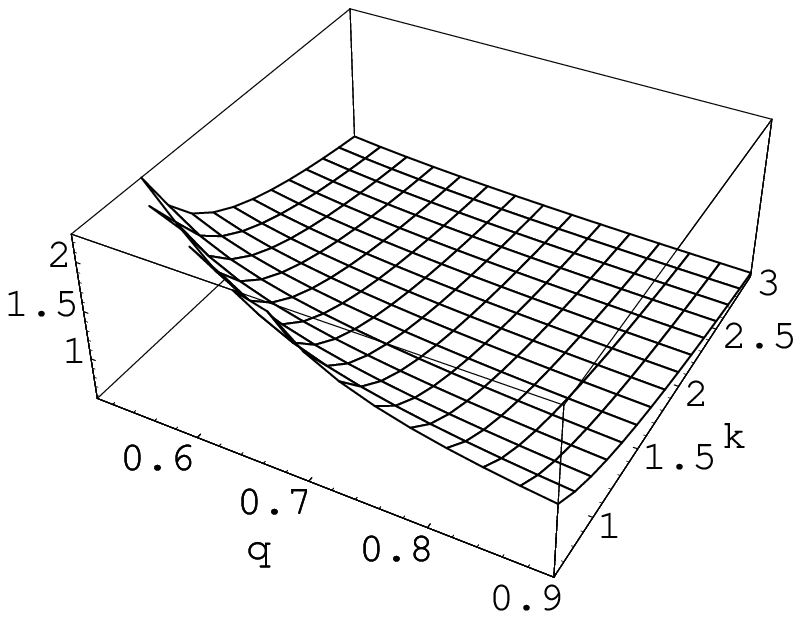, height=5cm}
 \end{minipage}
\begin{minipage}{.55\linewidth}
 \centering\epsfig{figure=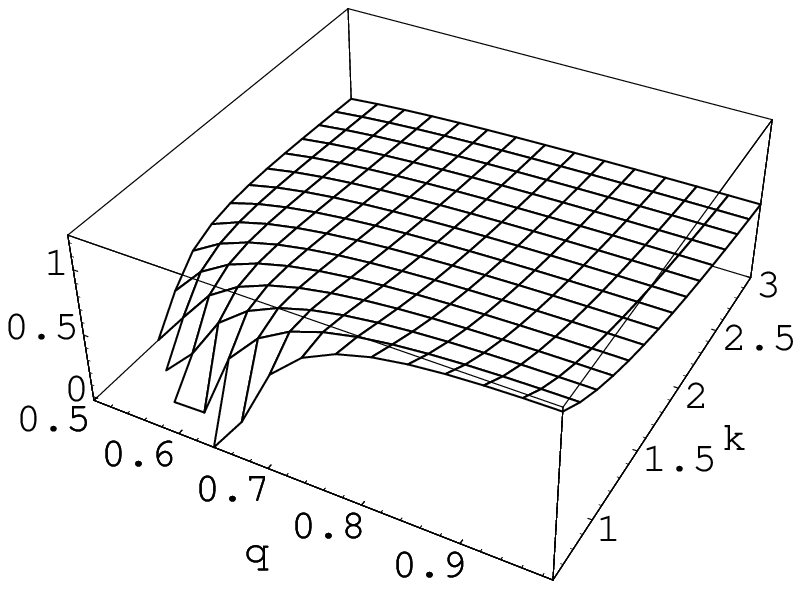, height=5cm}
  \end{minipage}
\caption{The entropies $S_R[p^R(x|q,\kappa)]/k_B$ (left) and
$S_B[p^R(x|q,\kappa)]/k_B$ (right) for power--law Hamiltonian with
the exponent $\kappa$ within the range $q>1/(1+\kappa)$.}
\end{figure}

Here, a useful extension of MEP is proposed. It consists in
looking for a maximum of the Renyi or Boltzmann entrpopies in
space of the Renyi distributions with different values of $q$.

It have appeared that both $S_R[p^R(x|q,\kappa)]$ and
$S_B[p^R(x|q,\kappa)]$ attain their maximums at boundaries of the
the range of possible values of $q$ defined by inequalities (15)
(see Fig. 1). Moreover, the Boltzmann entropy attains its maximum
value at $q=1$, that corresponds to the Gibbs distribution.

On contrary, the Renyi entropy attains its maximum value at the
minimal possible value of $q$ which satisfy  the inequality (15),
that is,
\begin{equation}
q_{min}=\frac1{1+\kappa}.
\end{equation}
 For $q<q_{min}$, the integral (10)
diverges and, therefore, the Renyi distribution does not determine
the average value $U=\langle H\rangle_p$, that is a violation of
the condition of MEP.

Thus, it is found that the maximum of the Renyi entropy is
realized at $q=q_{min}$ and it is just the value of the Renyi
parameter that should be used for the discussed particular case of
the power-law Hamiltonian if we have no additional information on
behaviour of the stochastic process under consideration.

Substitution of $q=q_{min}$ into Eq. (22) leads to
\begin{equation}
p\sim x^{-(\kappa+1)}
\end{equation}
Thus,  for $q=q_{min}$ the Renyi distribution for a system with
the power--law Hamiltonian becomes PLD over the whole range of
$x$.

For a particular case of the impact fragmentation where $H\sim
m^{2/3}$ the power-law distribution of fragments over their masses
$m$ follows from (21) as $p(m)\sim m^{5/3}$ that coincides with
results of our previous analysis \cite{fragm} and experimental
observations \cite{Fujiw}.

For another particular case $\kappa =1$  PLD is $p\sim x^2$. Such
form of the Zipf-Pareto law is the most useful in social,
biological and humanitarian sciences. Just the same exponent of
PLD was demonstrated \cite{Rybcz} in cosmic ray physics for energy
spectra of particles from atmospheric cascades.

It is necessary to notice here that inequalities (15) suggest in
fact $q>q_{min}$, that is,  $q=q_{min}+\epsilon$ where $\epsilon$
is a positive infinitesimal value. It is clear that $\epsilon (\ll
1)$ should be a finite constant in real physical systems.
Accounting for $\epsilon$ gives rise to the Renyi distribution in
the form
\begin{equation}
p^R(x)= Z^{-1}(C_u x)^{-(\kappa+1)(1 +\epsilon\frac
{\kappa+1}{\kappa})}\left[1-\epsilon(\kappa
+1)^2(1-C_ux^{-\kappa})\right]^{-\frac {\kappa+1}{\kappa}(1
+\epsilon\frac {\kappa+1}{\kappa})}
\end{equation}
For sufficiently great $x$'s  this Renyi distribution passes to
PLD where all terms with $\epsilon$ can be neglected.
\begin{figure}[t]
\begin{minipage}{.50\linewidth}
 \centering\epsfig{figure=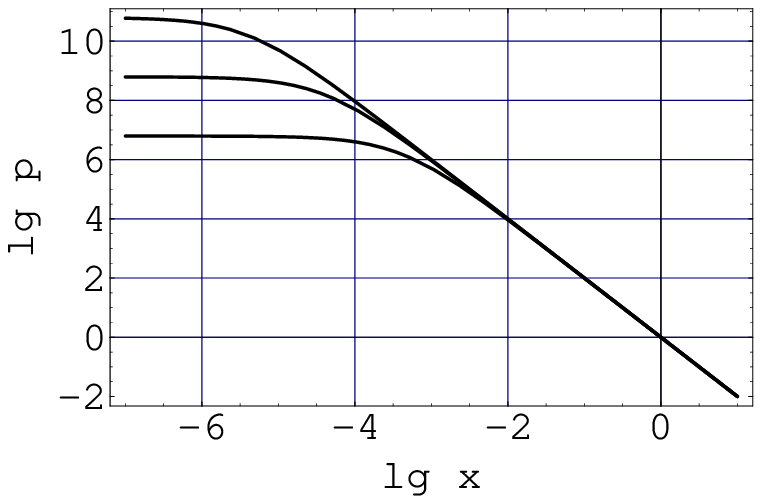, height=5cm}
 \end{minipage}
\begin{minipage}{.55\linewidth}
 \centering\epsfig{figure=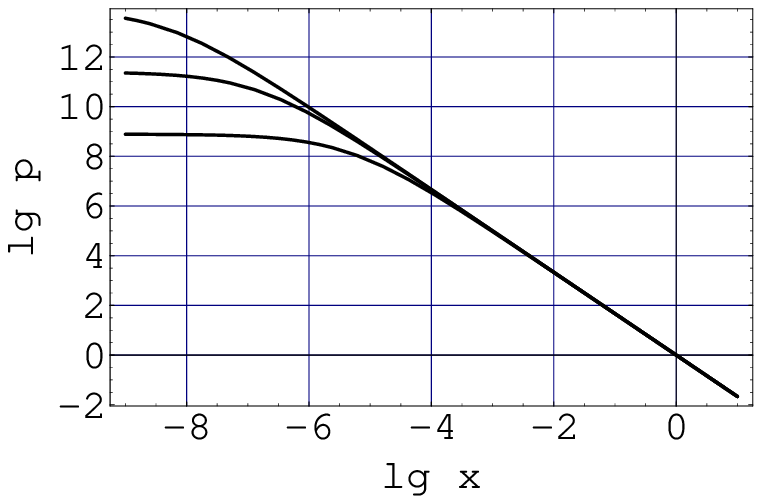, height=5cm}
  \end{minipage}
\caption{The Renyi distributions (non-normalized) for the
power--law Hamiltonian $H\sim x^\kappa$, $\kappa=1$ (left) and
$\kappa=2/3$ (right) and different values
$\epsilon=10^{-6},\,10^{-5},\,10^{-4}$ from upper to below in each
graph.}
\end{figure}

On the other side, for small $x\ll 1$, only the term
$\epsilon(\kappa +1)^2 C_ux^{-\kappa}$ may be accounted in the
expression in the square brackets, so we get
\begin{equation}
p^R(x)|_{x\ll 1}\sim(\epsilon(\kappa +1)^2)^{-\frac
{\kappa+1}{\kappa}}
\end{equation}
This equation points to the fact that the asymptote to the Renyi
distribution for small $x's$  is a constant of which value is
determined by  $\epsilon\ll 1$.

The picture of the Renyi distribution over the whole range of $x$
is illustrated in the Fig. 2 for particular cases of
$\epsilon=10^{-6}$, $\epsilon=10^{-5}$ and $\epsilon=10^{-4}$.

Now there is no methods for an unique theoretical determination of
$\epsilon$, so it may be considered as a free parameter. It can be
estimated for those experimental data where the head part preceded
PLD is presented. As an example, for the probability distribution
of connections in WWW network \cite{Wilk2} where the exponent of
PLD is equal to $-2.5$, the parameter $\epsilon$ is estimated as
$\sim 10^{-4}$.

\section{Conclusions}
Below, some results obtained in the present effort are summarized
briefly. Maximum Entropy Principle applied to the Renyi entropy
gives rise to a Renyi distribution that depends on the Renyi
parameter $q$ and two Lagrange multipliers $\alpha$ and $\beta$.
The multiplier $\alpha$ corresponds to the condition of
normalization of the distribution and may be eliminated with ease.
The second Lagrange multiplier $\beta$ corresponds to the
condition of a fixed average energy $U=\langle H\rangle_p$ just as
$\beta_0=1/k_BT$ in the Gibbs distribution function. It should be
noted here that the connection of $\beta_0$ with the thermodynamic
temperature obviates the necessity to eliminate the second
Lagrange multiplier from the Gibbs distribution. It is not so for
the Lagrange multiplier $\beta$, at least up to a time when the
new Renyi thermostatistics will be constructed and $\beta$ will
obtain a physical meaning.

It is shown here that for the particular case of a power--law
Hamiltonian $H_i=Cx^\kappa$ the Lagrange multiplier $\beta$ does
not depend on the Renyi parameter $q$ and coincides with
$\beta_0$. Moreover, it can be expressed in terms of $U$ and
$\kappa$ and thus excluded at all from the Renyi distribution
function.

In the absence of any additional information on a nature of the
stochastic process, the $q$-parameter of the corresponding Renyi
entropy is determined with the further use of MEP in the space of
$q$-dependent Renyi distributions. When applying such MEP to the
Boltzmann entropy the maximum is found at $q=1$ that corresponds
to the Gibbs distribution.  Maximum of  the Renyi entropy is found
at $q=1/(1+\kappa)$. The exponent of the power--law distribution
for such $q$ is $-(1+\kappa)$ that agrees with observable data for
stochastic systems with the power--law Hamiltonians.

It should be noticed that all above estimations of the parameters
of the Renyi distribution (7) and the exponent of PLD are true as
well for the  escort version of the Tsallis' distribution
\cite{Tsall}
\begin{equation}
p_i^{(Ts)}=Z_{Ts}^{-1} \left(1-\beta^*(1-q')\Delta
H_i\right)^{\frac{q'}{1-q'}}\\
\end{equation}
because of both distributions are identical if $q'=1/q$. Really in
this case
\begin{equation}
1-q'=\frac{q-1}{q}\,,\,\,\,\, \frac{q'}{1-q'}=\frac{1}{q-1}
\end{equation}
and $\beta^*$ is determined by the same second additional
condition of MEP as well as $\beta$.

\subsection*{Acknowledgements}
I acknowledge fruitful discussions of the subject with  A.
Vityazev.

\end{document}